\documentclass[amstex,aps,showpacs,preprint]{revtex4}%
\usepackage{graphicx}
\usepackage{amsmath}
\usepackage{bm}%
\usepackage{amsfonts}%
\usepackage{amssymb}

\begin{document}

\title{Exact quantization of nonsolvable potentials: the role of the
quantum phase beyond the semiclassical approximation}

\author {A.\ Matzkin}
\affiliation{Laboratoire de Spectrom\'{e}trie physique (CNRS
Unit\'{e} 5588), Universit\'{e} Joseph-Fourier Grenoble-I, BP 87,
38402 Saint-Martin, France}

\begin{abstract}
Semiclassical quantization is exact only for the so called
\emph{solvable} potentials, such as the harmonic oscillator. In
the \emph{nonsolvable} case the semiclassical phase, given by a
series in $\hbar$, yields more or less approximate results and
eventually diverges due to the asymptotic nature of the expansion.
A quantum phase is derived to bypass these shortcomings. It
achieves exact quantization of nonsolvable potentials and allows
to obtain the quantum wavefunction while locally approaching the
best pre-divergent semiclassical expansion. An iterative procedure
allowing to implement practical calculations with a modest
computational cost is also given. The theory is illustrated on two
examples for which the limitations of the semiclassical approach
were recently highlighted: cold atomic collisions and anharmonic
oscillators in the nonperturbative regime.

\end{abstract}

\maketitle

%PACS 32.60.+i 03.65.Sq 32.55.Be 05.45.Mt
%33.55.Be stark and zeeman for MOLECULES
%32.60.+i stark and zeeman for ATOMS

The semiclassical treatment of integrable systems, which can be
traced back to Bohr's atomic model of planetary motion and
Einstein's classic paper on the quantization of regular motion
\cite{bohr-einstein} is assumed to be a well-established and
venerable subject. It is true that WKB theory, where the phase to
be quantized is the classical action, can be found in any standard
textbook, but WKB often results in approximations that are
quantitatively too crude and that fail to capture the physics of
the problem. Indeed, WKB theory achieves exact quantization for a
restricted number of potentials - known as the solvable potentials
- such as the harmonic and Morse oscillators or the centrifugal
Coulomb problem.\ In the last decade, the application of
supersymmetric (SUSY) methods to quantum mechanics has enlarged
this list to a handful of other potentials, quantized by employing
SUSY WKB \cite{sukhatme97}. However, even in the solvable cases,
the WKB wavefunctions are innaccurate especially at the turning
points where they blow up, and a consistent divergence-free WKB
scheme is still a topic of current investigation
\cite{divfree2002}. In the more general nonsolvable case, WKB
quantization has frequently resulted in useful approximations to
compute the energy levels of excited states, but recently several
shortcomings were pointed out: for example the phase loss in the
classically forbidden regions is badly taken into account by WKB
theory in potentials used in atomic clusters calculations
\cite{friedrich96}; in cold atom collisions WKB quantization
breaks down for very excited states \cite{boisseau98}; for
anharmonic potentials, the failure of WKB has prompted extensive
developments of numerically involved quantum techniques with the
aim of obtaining accurate results \cite{hatsuda97,meissner97} for
applications ranging from molecular physics to quantum field
theories.

We show in this work that these shortcomings can be resolved by
considering an exact quantum phase. Although the quantum phase
bears a very close relationship to the semiclassical one, it is
necessary to go beyond the semiclassical approach in the presence
of nonsolvable potentials. Indeed, the failure of the
semiclassical approximation resides in the limitations imposed by
the underlying classical dynamics. Let us take a one dimensional
conservative system at energy $E$, which will be our main concern
in this work. The WKB wavefunction is built from blocks given by
\cite{brack badhuri}
\begin{equation}
\psi_{\mathrm{WKB}}(x,E)=A(x,E)\exp[iS(x,E)/\hbar] \label{e1}%
\end{equation}
where the phase function $S(x,E)$ is the classical action and $A^{2}%
(x,E)=(\partial_{x}S)^{-1}$ gives the classical probability
amplitude. The WKB
quantization condition reads%
\begin{equation}
S(t_{2},E)-S(t_{1},E)=\left(  n+\frac{\mu}{4}\right)  \pi\hbar, \label{e2}%
\end{equation}
where $t_{1,2}$ are the turning points, $\mu$ is the Maslov index
(generally 2 in one dimensional systems) and $n$ is the level
integer. To improve the approximation a semiclassical expansion
going beyond these purely classical terms can be carried out by
going to higher order in $\hbar$. However this expansion is an
asymptotic series, which means that although going to higher order
may improve the accuracy of the results, at some point the series
generally diverge \cite{bender77-voros93}.\ Moreover the
divergence of the amplitude at the turning points becomes worse at
each order and regularization techniques are needed to compute the
phase integrals.

This is why a quantum phase $\sigma(x,E)$ while implicitly summing
the divergent semiclassical expansion needs to be defined from the
start by appropriately transforming the Schr\"{o}dinger equation
\begin{equation}
\hbar^{2}\partial_{x}^{2}y(x)+p^{2}(x)y(x)=0,\label{e5}%
\end{equation}
where $p(x)$ is the classical momentum. A transformation of the
Liouville-Green type \cite{slavyanov96} is taken by writing
\begin{equation}
y(x)=\left(  \partial_{x}\xi(x)\right)  ^{-1/2}w(\xi),\label{e6}%
\end{equation}
so that $\xi(x)$ appears as a 'phase' and the prefactor as an
amplitude, as in Eq. (\ref{e1}).
Assume $w(\xi)$ fulfills the equation%
\begin{equation}
\hbar^{2}\partial_{\xi}^{2}w(\xi)+R(\xi)w(\xi)=0,\label{e7}%
\end{equation}
where the choice of the unspecified function $R(\xi)$ determines
the choice of $w$. $\xi$ then obeys
\begin{equation}
R(\xi)\left(  \partial_{x}\xi\right)  ^{2}-p^{2}(x)+\frac{\hbar^{2}}%
{2}\left\langle \xi;x\right\rangle =0,\label{e8}%
\end{equation}
where $\left\langle \xi;x\right\rangle
\equiv\partial_{x}^{3}\xi/\partial
_{x}\xi-\frac{3}{2}(\partial_{x}^{2}\xi/\partial_{x}\xi)^{2}$
denotes the Schwartzian derivative. We thus see that the choice of
the phase first depends on the choice of the carrier function $w$
(or equivalently, of $R(\xi)$), and then on the choice of the
boundary conditions that need to be imposed on the third order
nonlinear Eq. (\ref{e8}). This is an illustration of the ambiguity
suffered by phase functions in quantum mechanics, due here to the
fact that there is no unique manner to cut a given wavefunction
into a phase $\xi$ on the one hand, and an amplitude function
obeying the continuity equation
$\alpha(x,E)=(\partial_{x}\xi)^{-1/2}$ (as in the semiclassical
case) on the other.

Formally $\xi(x)$ can be expanded as an asymptotic series in
$\hbar$ irrespective of the specific choice of $R$. It is
nevertheless apparent from Eq. (\ref{e8}) that the only choice
that will give $\xi(x)\sim S(x)$ to first order in $\hbar$
corresponds to $R(\xi)=\pm1$, leading by Eq. (\ref{e7}) to
circular or exponential carrier functions. The price to pay is
that the $\hbar$ expansion based on these functions -- such as the
WKB approximation, necessarily diverges at the turning points.
However the exact solution of Eq. (\ref{e8}) does not present such
deficiencies, pointing to the possibility of defining a quantum
phase from Eq. (\ref{e8}) with $R(\xi)=\pm 1$. For bound state
problems the solutions of Eq. (\ref{e5}) are real so $w$ can be
taken as a circular function. $y(x)$ then becomes proportional to
$\alpha (x)\sin\sigma(x),$ with $\sigma$ denoting the phase $\xi$
when $R=\pm1$. Let $p(x)$ span the interval $]s_{1},s_{2}[$
(typically $]-\infty ,+\infty[$ or $]0 , +\infty[$ for radial
problems). It can then be shown \cite{matzkin01} that $\alpha$ is
a positive definite quadratic form that behaves as
$\alpha(x\rightarrow s_{i=1,2})\rightarrow\infty$
so that setting $\sigma(s_{1})=0$ the quantization condition reads%
\begin{equation}
\sigma(s_{2},E)=(n+1)\pi,\label{e9}%
\end{equation}
where $n$ is the level integer as in Eq. (\ref{e2}). Eq.
(\ref{e9}) is exact and holds irrespective of the boundary
conditions imposed on $\sigma$.

For computational purposes, Eq. (\ref{e8}) is intricate to solve.\
It can be checked that by writing (we use atomic units and assume
$R(\xi)=1$ from now
on)%
\begin{equation}
M(x,E)=\partial_{x}\left[  \sigma(x,E)+\frac{i}{2}\ln(\partial_{x}%
\sigma)\right]  ,\label{e10}%
\end{equation}
Eq. (\ref{e8}) leads to the complex but first order nonlinear
differential
equation%
\begin{equation}
\partial_{x}M=i\left(  p^{2}(x)-M^{2}(x)\right)  \equiv\mathcal{F}%
(M(x),x),\label{e12}%
\end{equation}
where $\mathcal{F}$ denotes the middle term taken as a functional.
Our strategy will consist in solving the equation for $M$; the
real part will give us $\partial_{x}\sigma,$ which can be
numerically integrated to obtain $\sigma,$ yielding both the
wavefunction and the total phase $\sigma(s_{2},E)$. To do so we
first linearize the equation for $M$ by expanding the functional
to first order in the vicinity of
an initial trial function $M_{0}(x)$. We then solve%
\begin{equation}
\partial_{x}M_{q+1}=\mathcal{F}(M_{q}(x),x)+\left.  \frac{\delta\mathcal{F}%
}{\delta M}\right|  _{M_{q}}\left(  M_{q+1}(x)-M_{q}(x)\right)  ,\label{e14}%
\end{equation}
with $q=0$ and where $\delta$ stands for the functional
derivative. Eq. (\ref{e14}) is a linear first order differential
equation that can be solved straighforwardly. Of course since
$\mathcal{F}$ has been linearized Eq. (\ref{e14}) is not
equivalent to Eq. (\ref{e12}), hence the subscript in $M_{1}$.
Convergence towards $M$ is achieved by iterating the procedure,
i.e. we now solve Eq. (\ref{e14}) for $q=1,$ obtaining a better
approximation $M_{2}$ and so on. This iterative linearization
procedure, known as the quasilinearization method (QLM) replaces
the solution of a nonlinear differential equation by iteratively
solving a linear one. It was introduced 3 decades ago by Bellman
and Kalaba \cite{bellman} in the context of linear programming,
but the extension of QLM to the type of functions dealt with in
quantum mechanics is quite recent \cite{mandelzweig99}. In
particular, the property of quadratic convergence which makes this
method powerful still holds.\ Indeed, 5 to 6 iterations typically
suffice to obtain solutions with high precision (e.g. more than 20
decimal digits).

To iteratively solve Eq. (\ref{e14}), two ingredients are needed:
first the trial function $M_{0}(x)$, second, the boundary
condition, $M_{q}(x_{b})$ which is set from the start since it
must be the same for each $q$. The choice of $M_{0}$ is actually
unimportant, since any sufficiently well behaved trial function,
i.e. behaving as $\left|  p(x)\right|  $ between the turning
points and as $i\left|  p(x)\right|  $ (resp. $-i\left|
p(x)\right| $) for $x<t_{1}$ (resp. $x>t_{2}$), with a smooth
transition between the regions, will lead to the same converged
answer. An elegant solution consists in determining the first
order semiclassical expression not diverging at the turning
points, which is done by appropriately choosing $R(\xi)$ in Eq.
(\ref{e7}) and expressing $\sigma$ in terms of $\xi_{R\neq1},$ but
we shall not pursue this task here \cite{matzkin-prep}. What is
crucial is the choice of the boundary condition because it
determines the behavior of the solution. We draw here on previous
work where we had shown that if $\sigma(x)$ is written as an
infinite $\hbar$ expansion, there is a unique boundary condition
that yields the classical action $S(x)$ when the limit
$\hbar\rightarrow0$ is taken \cite{matzkin01}. For other boundary
conditions, the limit $\hbar \rightarrow0$ yields a semiclassical
phase (and the corresponding amplitude) that displays the
oscillations of the WKB wavefunction. The existence of an optimal
boundary condition is well known in the context of solvable
potentials, where the use of analytic solutions (e.g. parabolic
cylinder functions for the harmonic oscillator, Whittaker
functions in the centrifugal Coulomb case \cite{seaton83}) allows
to explicitly construct a nonoscillating quantum phase. However
for nonsolvable potentials a procedure based on special functions
does not exist, whereas the $\hbar$ expansions employed in
\cite{matzkin01} are of formal nature. To ensure that the quantum
phase behaves as the $\hbar$ expansion would if it converged
\cite{arnold}, we choose here the simplest solution, namely to
pick an appropriate point $x_{b}$ in the classically allowed
region where the standard \ ($R(\xi )=1$) semiclassical expansion
can be employed and go to the highest possible order in $\hbar$
before the series starts to diverge. More refined methods of
estimating the optimal boundary condition, based on super and
hyper-asymptotic expansions \cite{boyd}, are in principle
available, but they are unnecessary insofar as neither the
behavior of $\sigma$ nor the quantum mechanical quantities would
be affected by their use.

\begin{figure}[tb]
\includegraphics[height=1.5in,width=2.1in]{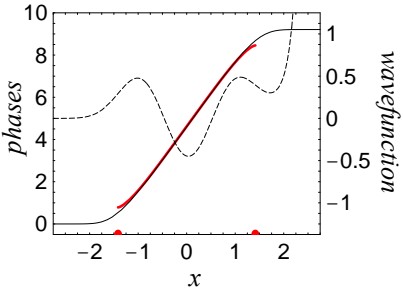}
\caption[]{Left scale: The quantum phase $\sigma(x)$ (solid black
curve)  for the anharmonic oscillator at $E=5$ (units in a.u.)
shown along with the classical action $S(x)$ (thick gray/red
curve) plotted between the turning points (semicircles on the $x$
axis). Note that we have taken $S(t_1)=\pi/4$. This $\pi/4$
constant is familiar from standard WKB analysis \cite{brack
badhuri} but it can also be rigorously proved
\cite{slavyanov96,matzkin-prep} that the semiclassical phase not
diverging at the turning points, obtained from Eq. (\ref{e8}) with
$R(\xi)=\xi$, behaves as $\sigma_{scl}(x) \backsimeq S(x) / \hbar
+ \pi/4$ for $x \gg t_{1}$ (but $\sigma_{scl}(t_1)=\pi /6$). Right
scale: The associated wavefunction $\alpha \sin \sigma$. Since $E$
is not an eigenvalue this improper wavefunction is normalized per
unit energy, as usual in scattering theory.\label{fig1}}
\end{figure}

We will now illustrate and detail the properties of the quantum
phase on two specific examples. The first example concerns
anharmonic oscillators which are employed in the investigation of
quite different phenomena, ranging from molecular vibrations to
quantum field theories and phase transitions.\ This has generated
different schemes to compute the eigenvalues and the
wavefunctions, calling for large numerical basis
\cite{macfarlane-annphys,meissner97} or delicate resummation
techniques \cite{hatsuda97}. The failure of the semiclassical
approximation which is important for the lowest levels (e.g., for
symmetric potentials $x^{2m}$ WKB quantization gives the wrong
behavior of the energies as a function of $m$) but persists for
more excited states, has been generally attributed to the
existence of complex turning points away from the real line
\cite{chebotarev99}.\ However the complex plane does not play
directly any r\^{o}le in the present scheme, based on the
construction of a real quantum phase. Fig.\ 1 shows the quantum
phase $\sigma(x,E)$ along with the classical action for a
Hamiltonian with an anharmonic potential $x^{2}+2x^{4}$.\ Given
the symmetry of the potential, we have taken $x_{b}=0$ and fixed
the value $M_{0}(x_{b})$ by carrying out a semiclassical expansion
up to $o(h^{14})$. We have also plotted on the same figure the
associated wavefunction. It diverges beyond the right turning
point because we have voluntarily chosen $E$ not to be an
eigenvalue (here $E$ lies between the 2nd and the 3rd levels). The
bound states are found by applying the quantization condition
(\ref{e9}) in the following way: the total phase $\sigma(s_{2},E)$
at $s_{2}=\infty$ is determined for different values of the energy
(the energy grid can be more or less sparse, depending on the
required numerical precision).\ The points are then interpolated
to yield the curve $\sigma(s_{2},E)$ as a function of $E$ seen on
Fig. 2, which allows to solve for $E$ in Eq. (\ref{e9}). By that
method we have determined the ground and the first 35 states of
this anharmonic oscillator and compared the eigenenergies to the
precise quantum calculations of \cite{meissner97}, obtaining
exactly the same results as those published in their Table III.
\begin{figure}[t]
\includegraphics[height=1.5in,width=2.1in]{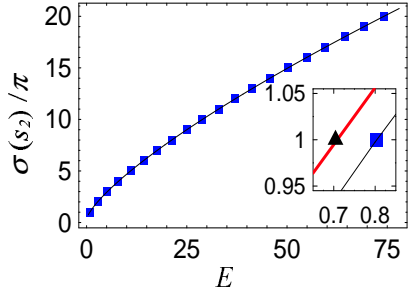}
\caption[]{Energy levels of the anharmonic oscillator (a.u.)
obtained by quantizing the quantum phase. The smooth curve
represents $\sigma(s_2,E) / \pi$ as a function of the energy and
the boxes are plotted each time this quantity equals the integer
$n+1$, determining the exact levels (here the first 20 levels
$n=0-19$ are plotted). The inset zooms on the ground state: the
thick gray/red line gives the semiclassical quantization curve
$(S(t_2,E) - S(t_1,E))/ \pi + 1/2$ (the $\hbar$ expansion diverges
beyond this order) and the triangle the semiclassically quantized
level, which is off by more than 10\%. \label{fig2}}
\end{figure}

The second illustration involves a potential well with a strong
repulsion at short-range and a long-range attractive tail. This
type of potential is of interest in the study of cold atomic
collisions, a field that has been sparked by the development of
photoassociative spectroscopy \cite{boisseau98}.\ The WKB
approximation breaks down for excited states near the threshhold,
a fact that is hardly surprising since it is understood that WKB
may fail when the quantum particle explores large areas of
classically forbidden regions\cite{friedrich04}. We take the
following 12-6 Lennard-Jones (LJ)
potential with the classical momentum given in the scaled form as%
\begin{equation}
p^{2}(x)=B\left[  E-\left( \frac{1}{x^{12}}-\frac{2}{x^{6}}\right)
\right] \label{e30}
\end{equation}
where $B$ is a 'strength' parameter. This potential with
$B=10^{4}$ has often been employed as a benchmark (see
\cite{leroy83-boisseau00,friedrich04} and Refs. therein) and is
known to support 24 states. The WKB quantization condition
(\ref{e2}) gives energies with an error relative to the local
level spacing that globally increases with
$E$. Fig. 3 compares the derivative of the quantum phase $\partial_{x}%
\sigma(x,E)$ with the classical momentum $\partial_{x}S(x,E)$ for
the last bound state just below the threshold. $x_{b}$ was taken
to the right of the potential minimum and the $\hbar$ expansion
was carried out up to the 12th order. The 2 curves are barely
distinguishable in the classically allowed region, so we have
focused on the zone near the turning points. Note that the quantum
curve largely penetrates into the classically forbidden region
well beyond the outer turning point. This is typical of excited
states in potentials with a long-range attractive tail and can be
seen as the underlying reason ruling the breakdown of the
semiclassical approximation. Quantization proceeds as in the first
example, by determining $\sigma(s_{2},E)$ on an energy grid then
solving for Eq. (\ref{e9}).\ The resulting curve is shown in Fig.\
4 and the quantized energies of the 24 levels exactly reproduce
the values of exact quantum mechanical calculations
\cite{friedrich04}. In particular the position of the last level
determines the scattering length, a crucial parameter in the
production of Bose-Einstein condensates.

\begin{figure}[tb]
\includegraphics[height=1.5in,width=2.1in]{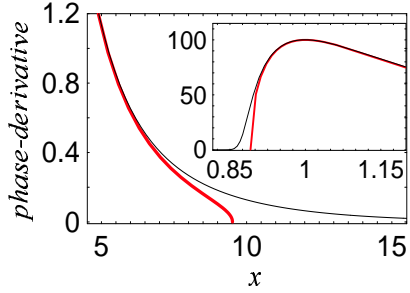}
\caption[]{Derivative of the quantum phase $\partial_{x} \sigma$
(solid black curve) and of the classical action $\partial_{x} S=
p$ (thick gray/red curve) in the 12-6 LJ potential characteristic
of cold atom collisions (scaled a. u.). These quantities, plotted
for the most excited bound state, are shown in the zone near the
outer turning point $t_2\thickapprox 9.5$. The inset shows the
zone near the inner turning point $t_1\thickapprox 0.9$ (note the
very different scales). \label{fig3}}
\end{figure}

To summarize, we have employed a quantum phase that goes beyond
the semiclassical approximation in giving the exact quantization
energies as well as the wavefunctions. We have also given a
numerical procedure to determine the phase based on the
linearization of the phase equation and a boundary condition
obtained from a local semiclassical expansion, and illustrated the
approach in the case of two nonsolvable potentials.

\begin{figure}[t]
\includegraphics[height=1.5in,width=2.1in]{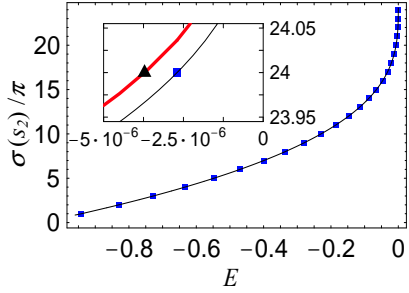}
\caption[]{Exact energy levels (boxes) of the 12-6 LJ potential
obtained by quantizing $\sigma (s_2,E)$ (solid curve), in scaled
a.u. The inset zooms on the last state below threshold ($E=0$) and
also shows the WKB prediction, about 40\% too low (triangle).
\label{fig4}}
\end{figure}


\begin{thebibliography}{9}
\bibitem{arnold}As put by Arnold (in \emph{Mathematical Methods of Classical
Mechanics} (Springer, Berlin, 1989)), the divergence of the series
means that 'we are looking for an object that does not exist' as
an expansion.
\end{thebibliography}
\end{document}